\documentclass[twocolumn,aps,showpacs,preprintnumbers,amsmath,amssymb,amsfonts,showkeys]{revtex4}

\usepackage{graphicx}
\usepackage{epsfig}
\usepackage{dcolumn}
\usepackage{bm}
\usepackage{latexsym}

\begin{document}

\title{Synthesize Neutron-Drip-Line-Nuclides with Free-Neutron Bose-Einstein Condensates Experimentally}

\author{ Bao-Guo \,Dong }
\affiliation{\\
Department of Nuclear Physics, China Institute of Atomic Energy,
P.O. Box 275 (10), Beijing 102413, China }
\date{\today}

\begin{abstract}
We first show a possible way to create a new type of matter,
free-neutron Bose-Einstein condensate by the ultracold
free-neutron-pair Bose-Einstein condensation and then determine the
neutron drip line experimentally. The Bose-Einstein condensation of
bosonic and fermionic atoms in atomic gases was performed
experimentally and predicted theoretically early. Neutrons are
similar to fermionic atoms. We found free neutrons could be cooled
to ultracold neutrons with very low energy by other colder neutral
atoms which are cooled by the laser. These neutrons form neutron
pairs with spin zero, and then ultracold neutron-pairs form
Bose-Einstein condensate. Our results demonstrate how these
condensates can react with accelerated ion beams at different energy
to synthesize very neutron-rich nuclides near, on or/and beyond the
neutron drip line, to determine the neutron drip line and whether
there are long-life nuclide or isomer islands beyond the neutron
drip line experimentally. Otherwise, these experimental results will confirm our prediction that is
in the whole interacting region or distance of nuclear force in all energy region from zero to infinite, Only repulsive nuclear force exists among identical nucleons and only among different nucleons exists attractive nuclear force.
\end{abstract}



\maketitle


The Bose-Einstein condensation of bosonic and fermionic atoms in
atomic gases was performed experimentally
\cite{And95,Dav95,Reg03,Gre03} and predicted theoretically early
\cite{Ein25}. Coupling fermionic atoms to bosonic molecules, thus
this alters the quantum statistics of the atomic Fermi gas system.
In the next level of matter structure, there is a neutral fermionic
particle, neutron, which is similar to fermionic atoms. However,
there is no object formed only with neutrons in a vast mass range of
covering from a single neutron up to neutron stars on observations.
Here we first show a possible way to create a new type of matter,
free-neutron Bose-Einstein condensate by the ultracold
free-neutron-pair Bose-Einstein condensation and then determine the
neutron drip line experimentally. We found free neutrons could be
cooled to ultracold neutrons with very low energy by other colder
neutral atoms which are cooled by the laser. These neutrons, trapped
in magnetic trap, form neutron pairs with spin zero, and then
ultracold neutron-pairs form Bose-Einstein condensate. Our results
demonstrate how these condensates can react with accelerated ion
beams at different energy to synthesize very neutron-rich nuclides
near, on or/and beyond the neutron drip line, to determine the
neutron drip line and whether there are long-life nuclide or isomer
islands beyond the neutron drip line experimentally. We anticipate
our assay to be a starting point for more sophisticated researches
in the free-neutron Bose-Einstein condensate. For example, free
neutrons are unstable and neutrons bound in deuteron are stable,
then what about neutrons in the free-neutron Bose-Einstein
condensate are? In which the interaction strength is between strong
interaction and free, the extremely weakly binding. In the future
there are many such interesting features of the free-neutron
Bose-Einstein condensates will be demonstrated and expected.

Free neutrons are unstable particles which can decay into a proton,
electron and neutrino, and with life-time about 896 s and spin 1/2
$\hbar$ \cite{Hu87}. The energy of free neutrons covers a vast range
from 10$^{-7}$ to 10$^9$ eV by different neutron sources which
include radioactive nuclide, accelerator, spallation, and reactor
neutron sources. The middle energy of neutrons produced by
$^{239}$Pu fission is 2 MeV. Here we only concentrate on
nucleon-nucleon interactions with energy of $E \sim 0$, and the
traditionally investigating energy range for nuclear structures is
$E \ll 0$, the bound states, and nuclear reactions $E \gg 0$, of the
energy order of MeV.

The nucleon-nucleon interaction or nuclear force is not full clear
to understand by now due to its complicacy and is dependent on
relative distance, relative movement situation, relative momentum of
interacting nucleons, spin, isospin, energy, density etc. There
exists no derivation of the nucleon-nucleon force from first
principles. So it is difficult to confirm the features of nuclear
force by the existing theoretical nucleon-nucleon interactions, such
as the bare nucleon-nucleon forces, microscopic effective
interactions, and phenomenological effective interactions,
especially under the extreme conditions at low energy of $E \sim 0$
and near the interacting boundary of nuclear force. The main
features of nuclear force are short range and saturation character.
By the model theories supported by experiments, when two nucleons
are near and interact with strong interaction, the main attractive
effect comes from the s-wave, and p-wave for repulsive. For two
different nucleons, obey the Pauli principle and total wave function
asymmetry under the two particle exchange, the spin direction can be
parallel or antiparallel, and the attractive interaction strength of
parallel is even larger than the antiparallel. For two same
nucleons, fermion and identical particle, only the spin antiparallel
state can exist the relative movement s-state, which decreases the
probability in the s-state and weakens the interaction \cite{Hu87}.
For these the experimental evidence is deuterium, $^2$H, exists and
no $^2$He.

In general, if an exchange of the coordinates causes the change of
the space coordinate part of wave function is symmetry there is an
attractive force between the two nucleons, and if asymmetry the
force is repulsive. For two neutrons the spin part should be
asymmetry for spin antiparallel case and the space coordinate part
should be symmetry, that is, the force is weak attractive between
the two spin antiparallel neutrons, and could much weaker than
deuterium. The long distance part of nuclear force is attractive and
the short part is repulsive, which is confirmed and demonstrated by
experiments \cite{Hu87}. So when two ultracold neutrons with very
low energy of lower or about $10^{-7}$ eV meet, we can  expect to
achieve these two neutrons to form a loosely bound pair of neutrons
with spin zero, i.e. a boson, or form it under some certain conditions which
are similar to fermionic gases. Many neutron pairs could form the
free-neutron-pair Bose-Einstein condensate when they are cooled
further and/or treated as fermionic atoms. We think the nuclear
theory, at least in the $10^{-7}$ eV energy region and
the interacting boundary region of nuclear force,
is not good enough to confirm this prediction or conclusion
but experiments could confirm it. The properties of nuclear force
are unclear by now, especially in the interacting boundary region,
which could be investigated experimentally in this way.
Experimenters can accurately test features of low energy pp
collisions first for easy detectability in experiments.

There are many cooling methods to cool free neutrons to produce cold
neutrons at energy of $10^{-3}$ eV, such as liquid hydrogen, H$_2$,
or liquid deuterium D$_2$ cold neutron sources in reactors. We can
obtain ultracold neutrons with energy on the order of $10^{-7}$ eV
from cold neutrons directly by the traditional methods, such as
vertical or curved horizontal channel extraction from cold neutron
sources in reactors, or by the Steyerl turbine, beryllium converter,
superfluid helium, or solid deuterium \cite{Ign96}.

Here we suggest a new effective and suitable method, by atom
indirect cooling technology, to cool the cold neutrons with the
colder medium atoms to the super low temperature. We can select
appropriate atom, such as hydrogen, lithium, sodium, or rubidium, as
the cooling medium atoms, adopt the laser cooling technology to cool
the atoms and bound in the magnetooptic trap. Then we cool the cold
neutrons by the colder medium atoms to the ultracold neutrons at
energy of $10^{-7}$ eV or lower, which becomes the ultracold neutron
gas. These further cooling neutrons are collected into the magnetic
trap. Now experimenters can trap ultracold neutrons by the magnetic
trap by experimental technique, then they can gradually cool them to
form super ultracold neutrons when need.

We have to develop suitable method and technology to
increase cold and ultracold neutron density. One possible method is
the multi-grade magnetic compression method, which may be adopted to
increase the ultracold neutron density. For further cooling the
ultracold neutron temperature or decreasing their energy, similar to
the atomic cooling method, by decreasing the binding potential trap
depth or radio frequency resonance reverse bound angular momentum or
spin states to non-bound angular momentum states induced evaporative
cooling technology, we force or select higher energy atoms and
neutrons to evaporate to make the rest neutrons colder reach to or
below critical temperature or Fermi temperature, and become the high
degenerate quantum Fermi gas, and the high density ultracold neutron
gas.

The high density ultracold neutrons could be directly used to do
nuclear physics researches, such as design as the high density
ultracold neutron gaseous target to synthesize neutron-rich nuclei
as well as nuclei near the neutron drip line experimentally, and for
their ground state property research, or the high accuracy
measurement of neutron charge, neutron mean lifetime and neutron
electric dipole moment, which are very important basic physical
quantity in verification of basic theoretical predictions.

 The extremely difficult challenge of further development is to form
the neutron pairs. Due to free neutrons without bound state and very
weak magnetic interaction, we have to seek and find the proper
method and mechanism to form the neutron pairs, like the creation of
bosonic molecules from fermionic $^{40}$K atoms using a magnetic
field ramp across a Feshbach resonance \cite{Reg03}, make high
degenerate neutron gas form free-neutron pair gas, namely superfluid
gas or gaseous superfluid. One possible mechanism or way to form the
pairs is by the strong interaction between two neutrons near the
interacting boundary of nuclear force. Other possible mechanisms to
form the pairs are by the magnetic interaction, as fermionic atom
condensations, strengthened by magnetic resonances due to magnetic
moment of neutrons.

The challenges include that one has to make the low neutron density
high enough to enhance the interaction probability to form pairs
with spin zero. This needs to investigate in physics, technology and
corresponding equipments.

Ultimately, these neutron pairs form free-neutron-pair Bose-Einstein
condensate which performs free-neutron Bose-Einstein condensation
process. The detecting methods of the temperature of ultracold
neutrons and the existence of the free-neutron Bose-Einstein
condensate are difficult for experiments. The laser fluorescence
detected method used in atoms is not suitable since neutrons do not
absorb a photon to excite to the excited state, such as the radio
frequency pulse fermionic molecule dissociation method \cite{Gre03}.
We can think that the temperature of ultra neutrons is the same as
that of the cooling medium atoms. To determine the temperature of
further evaporative cooling neutrons and the condensates is an open
problem. The possible characteristic photon emitted by the coupling
of two ultracold free-neutrons could be as the detected signal for
the pair formed.

\begin{figure*}
\includegraphics[width=8.6 cm]{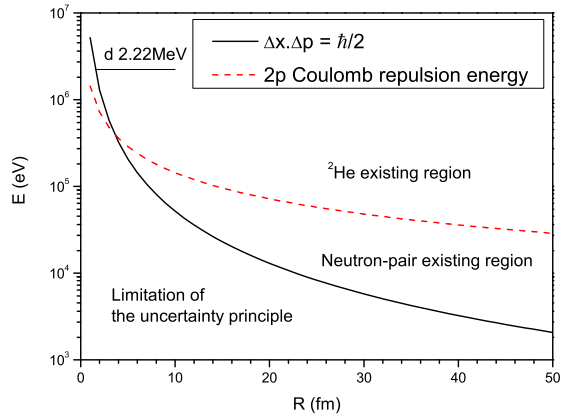}
\caption{ (Color online) Limitation of the uncertainty principle,
$\Delta x \cdot \Delta p \geq \hbar/2$, to the pair formed. Above
the line is the possible binding state region and below the
non-binding region. The possible binding state region including Coulomb repulsion of $^2$He is shown.}
\label{fbc}
\end{figure*}

The limitation of first principles have been considered carefully.
The limitation of the uncertainty principle, $\Delta x \cdot \Delta
p \geq \hbar/2$, to the pair formed is shown in Fig. \ref{fbc}. This
principle gives the possible distance between the two interacting
neutrons and the minimum binding energy they have to need if they
form a pair with spin zero. The binding energy of deuterium is
2.2246 MeV so the nearest distance for the two nucleons is 1.526 fm
under the limitation of the uncertainty principle. When the distance
is 10 fm the binding energy needed for two neutrons decreases to 52 keV, see Fig.
\ref{fbc}.

Once the free-neutron Bose-Einstein condensates are created
experimentally, it can be as a target to react with accelerated ion
beams at different energy to synthesize very neutron-rich nuclides
near the neutron drip line, especially heavy nuclides on and beyond
the neutron drip line, to determine the neutron drip line and
whether there are long-life nuclide or isomer islands beyond the
neutron drip line experimentally. Since there is no Coulomb barrier,
the energy of neutrons in the condensates is ultra low and the
neutron number density could be the highest artificially
experimentally usable density, so we anticipate this could
synthesize the most neutron-rich nuclides which are near, on, and
beyond the neutron drip line. Similarly, we can search after the
maximum neutron magic number in the isotopic chains which could
include super heavy elements or nuclides.

By this way, the free-neutron Bose-Einstein condensation, for
further development, the macroscopic quantitative neutron nuclear
matter could be created on artificiality. This can provide the
second experimental measurable nuclear matter density point for the
research of macroscopic nuclear matter properties experimentally,
and the verification of the theoretical models and their predicted
results. On the other hand, neutron mean lifetime and neutron
electric dipole moments can be measured with high accuracy.

In the future, experimental results will finally confirm the existence of weakly binding free-neutron-pair.
Otherwise, these results will confirm our other prediction:
In the whole interacting region or distance of nuclear force in all energy region from zero to infinite, only repulsive nuclear force exists among identical nucleons and only among different nucleons exists attractive nuclear force. That indicates among p+p or n+n only the repulsive nuclear force exists, and the attractive nuclear force only exists between p+n interaction.

In summary, we have suggested a new type of matter, free-neutron
Bose-Einstein condensate and its experimental creation way, a new
neutron cooling method, i.e. by other colder neutral atoms which are
cooled by the laser, the mechanism of free-neutron-pair formation,
i.e. possible way to form the pairs is by the strong interaction
between two neutrons near the interacting boundary of nuclear force,
 or by the magnetic interaction, potential
methods to synthesize heavy nuclides on the neutron drip line and to
determine the neutron drip line and whether there are long-life
nuclide or isomer islands beyond the neutron drip line
experimentally, possible way to produce macroscopic quantitative
man-made neutron condensed matter. That will open a new research
field, \emph{zero energy} nuclear physics.
Otherwise, these experimental results will confirm our prediction that is
in the whole interacting region or distance of nuclear force in all energy region from zero to infinite, Only repulsive nuclear force exists among identical nucleons and only among different nucleons exists attractive nuclear force.

This work is supported by the National Natural Science Foundation of
China under Grant No. 11075217.

\appendix Problems asked and answers responded

\begin{enumerate}
\item
\begin{description}
\item[Ask] In this manuscript the author suggests a possible way for creating a free-neutron Bose-Einstein condensate made of ultracold neutron pairs. I found this study as very descriptive and without content and in my view it should not be published.
\item[Answer] I do not agree your judgements for this manuscript because your basic assumptions are incorrect and you use the incorrect theory and methods, both in energy region and nuclear force, so you conclusions are unfair, incorrect and unreliable.

　　The manuscript this manuscript is based on first principles, not on approximate nuclear theory or models in which the accuracy is only about 1 MeV.

　　It is well known that in different energy region different theory or method should be used in nuclear physics. What you used is the general nuclear theory which is suitable in the 1-50 MeV energy region or higher. But in this manuscript the energy is about $10^{-7}$ eV. Here the difference is about 10-13 order of magnitude in the two situations. You used the same methods for such low energy incoming neutrons and must give the incorrect conclusions.

　　The other is nuclear force that you think true correction or truth. The real situation is the existing nuclear force can not describe the loosely binding halo nuclides near the drip line and not derive from first principles. It is unclearly now, especially in the boundary region of nuclear force. So we can not give the formula in this region and can not confirm it with the existing nuclear force. That is why we suggest to confirm it with experiments only.

　　By the comments given, without pointing out any actual incorrectness, not an suitable academic comment, I think you has not read the manuscript this manuscript seriously or carefully and understood it correctly. The reasons stated as above for no formula.

　　The manuscript this manuscript Predicts or suggests:
(1) a new type of matter, free-neutron Bose-Einstein condensate;
(2) its creation way;
(3) a new neutron cooling method: by other colder neutral atoms which are cooled by the laser;
(4) the mechanism of free-neutron-pair formation: One possible way to form the pairs is by the strong interaction between two neutrons near the interacting boundary of nuclear force; Other possible mechanisms to form the pairs are by the magnetic interaction strengthened by magnetic resonances due to magnetic moment of neutrons, or by the coherence of the de Broglie wave of　ultracold neutrons.
(5) potential methods to synthesize heavy nuclides on the neutron drip line and to determine the neutron drip line and whether there are long-life nuclide or isomer islands beyond the neutron drip line experimentally;
(6) possible way to produce macroscopic quantitative man-made neutron condensed matter.
　　Consider the list above, I think your comment "without content" is unfair and incorrect to this manuscript.

　　In summary I think the manuscript this manuscript has very high academic values because so many innovations are predicted or suggested.
\end{description}

\item
\begin{description}
\item[Ask] The proposed method to produce a Bose-Einstein condensate of free
neutrons is challenging and innovative, indeed. If it could be
actually realized it would open a spectacular new route in the quest
for islands of stability beyond the neutron drip line.

But what I miss is a more elaborate assessment of its experimental
feasibility which meets the high standards. A sentence (page 2)
like ＾This needs to investigate in physics, technology and
corresponding equipments.￣ is surely not enough.

Another reason why this short manuscript is not publishable in its
present form is its bad English, which requires massive revision,
preferably by a native speaker.

     \item[Answer] The manuscript this manuscript is based on first principles, not on approximate nuclear theory or models in which the accuracy is only about 1 MeV.

　　It is well known that in different energy region different theory or method should be used in nuclear physics. What you used is the general nuclear theory which is suitable in the 1-50 MeV energy region or higher. But in this manuscript the energy is about $10^{-7}$ eV. Here the difference is about 10-13 order of magnitude in the two situations. You used the same methods for such low energy incoming neutrons and must give the incorrect conclusions.

　　The other is nuclear force that you think true correction or truth. The real situation is the existing nuclear force can not describe the loosely binding halo nuclides near the drip line and not derive from first principles. It is unclearly now, especially in the boundary region of nuclear force. So we can not give the formula in this region and can not confirm it with the existing nuclear force. That is why we suggest to confirm it with experiments only.

　　You admitted the innovations in this manuscript, but hope to know a more elaborate assessment of its experimental feasibility. I do not think your requirement is reasonable. Because the extremely low energy region is not suitable for the existing nuclear theory, and the existing nuclear force is not suitable for the very loosely binding states, as stated above, so this assessment of its experimental feasibility can not be given or confirmed by theory. Showing a possible way with physics mechanism should be finished or performed by theory or theorist, the detailed experiments should be performed by experimentalists, I think.
\end{description}

\item
\begin{description}
\item[Ask] In this manuscript, the author proposes to create a Bose-Einstein
condensate of free dineutrons. A free dineutron is a hypothetical
bound particle formed by two free neutrons.

The manuscript displays a shocking lack of scientific quality in all
conceivable respects. It is completely unsuitable for publication in
any scientific journal.

The text is a poorly structured and poorly phrased narrative that is
incomprehensible at many points. Most statements do not go beyond the
level of random guessing. They are typically not backed up by
references to the literature. There is no backup coming from serious
calculations either, except for the triviality of Fig. 1. Most things
remain on a completely qualitative level. As far as I can see, there
is no new physics in the manuscript beyond the storytelling aspect of
tossing out unrealistic ideas at random.

I am not an expert on dineutrons but a brief literature search left me
with the impression that a free dineutron is not expected to exist and
has never been observed. There seems to be an ongoing debate whether
dineutrons exist within certain nuclei and whether they are involved
in certain nuclear decay processes. There may also be very short-lived
resonances in the scattering of two neutrons. But their short lifetime
makes them unsuitable for the proposal here. As long as there is no
realistic perspective, that (and why) this will change, proposing a
BEC of these hypothetical particles is completely pointless.

\item[Answer] I do not agree your comments on judgement for the scientific value of this manuscript because your assumptions is incorrect both in energy region and nuclear force, as stated above.

　　I think you try to explain and judge this manuscript in the frame of existing nuclear theory and you do not pay attention to whether its physical assumptions are suitable. What this manuscript shows is beyond the physical assumptions of existing nuclear theory, both in energy region and nuclear force, which is stated in the beginning of the manuscript this manuscript.

\end{description}

\item
\begin{description}
\item[Ask] There is plenty of literature on dineutrons. None of it is referenced.
The fact that free dineutrons have not been observed and are not
expected to exist is not even particularly clearly stated in the
manuscript.
\item[Answer] The key point is energy. Dineutron in literatures is at high energy of MeV of GeV, if you check, but here in this manuscript the energy is about $10^{-7}$ eV. So those literatures are not suitable here.
\end{description}

\item
\begin{description}
\item[Ask] The fact that a proton plus a neutron can form a bound state, the
deuteron, is not a sufficient reason to claim qualitatively that two
free neutrons would form a bound state. There are different numbers of
up and down quarks involved in the two cases and a serious calculation
would be needed to make any such claims.
\item[Answer] Deuteron is an example to show that two neutrons form a very loose binding boson not violating first principle. For serious calculations, you would like to suggest based on QCD or nuclear force / potential, but no nuclear force derived from first principles now so your suggestion is impossible to perform by the field, I think.
\end{description}

\item
\begin{description}
\item[Ask] Unlike alkali atoms, neutrons do not have ground-state hyperfine
structure. As a result, magnetically tunable Feshbach resonances like
the ones used to associate two ultracold fermionic atoms to one
bosonic molecule do not exist. The alternative methods for associating
dineutrons described in the manuscript are incomprehensible.
\item[Answer] That means originally to form a boson with very weak magnetic interaction between two neutrons, not by their ground-state property. Other methods can be confirmed by experiments only as stated above.
\end{description}

\item
\begin{description}
\item[Ask] If dineutrons existed at all, they would have a total spin of zero.
Hence, they do not display a linear Zeeman effect and it will be
difficult to trap them magnetically.
\item[Answer] There are many difficulty to perform the tasks in this manuscript for experiment, most of them are challenging, include this one. But it can be solved by further researches. Difficulty is not no high academic value, I think. Require one paper to solve all difficulty is not reasonable.
\end{description}

\item
\begin{description}
\item[Ask] If one wants to trap neutrons in a magnetic trap, they have to be spin
polarized because the magnetic trapping force changes sign when the
spin orientation changes. If two neutrons in a spin polarized gas
approach each other to hypothetically form a dineutron, they will be
in the triplet state. Associating neutrons in a magnetic trap will
therefore not be able to create spin-zero dineutrons.
\item[Answer] This problem has been solved in fermionic atom BEC experiment by changing static magnetic field into alternating field.
\end{description}

\item
\begin{description}
     \item[Ask] Sympathetic cooling of free neutrons (or free dineutrons) by spatially
overlapping them with an atomic gas that is being laser cooled will
only be possible if the collision between the atoms and the neutrons
are predominantly elastic and if the scattering cross section for
these collisions is high enough to achieve thermalization during the
lifetime of a neutron. There is no quantitative analysis discussing
what the scattering properties are.
\item[Answer] One way is to select suitable atomic gas as suggested in this manuscript. The scattering cross section at energy $10^{-3\sim-7}$ eV can not be calculated using the existing nuclear force or potential as stated above. This suggestion is good but can not be performed now because of no suitable nuclear potential at such low energy.
Even I do so, others would think it is not confirmed by any experiment and unreliable.
\end{description}

\item
\begin{description}
\item[Ask] I do not find sufficient merit in this paper to warrant publication. Many reasons were given by at least one reader to support a rejection, and I will not reiterate them here although I find them all valid.
     \item[Answer] The manuscript this manuscript is based on first principles, not on approximate nuclear theory or models in which the accuracy is only about 1 MeV.

　　Reader¨s conclusions or judgements only and show an unfair hearing manifestly. reader thinks them (i.e. reader¨s reasons) all valid, but without pointing out what these valid evidences are based on, should be understood the same as reader¨s. Here to review this manuscript, the three readers used models or theory based on the wrong basic assumptions both in energy region and nuclear force, i.e. 1-50 MeV energy region / $10^{-7}$ eV energy region and approximate nuclear force not derive from first principles, as I respond to the reader¨s comments.

　　It is well known that in different energy region different theory or method should be used in nuclear physics. What the readers used is the general nuclear theory which is suitable in the 1--50 MeV energy region or higher. But in this manuscript the energy is about $10^{-7}$ eV. Here the difference is about 10--13 order of magnitude in the two situations. They used the same methods for such low energy incoming neutrons and must give the incorrect conclusions. So reader¨s conclusions, the same as reader¨s, are incorrect and unfair.
\end{description}

\item
\begin{description}
\item[Ask] The bottom line is the fact that there is no bound state for two neutrons even at zero temperature.
     \item[Answer] This state is correct only at very limited conditions, not correct for all conditions. But reader omitted the most key point, the required condition. Here it is both energy region and nuclear force.

　　Because there is no experimental evidence and theory for its existing by now the author first predicts a possible way to create a new type of matter, free-neutron Bose-Einstein condensate by the ultracold free-neutron-pair Bose-Einstein condensation and then determine the neutron drip line experimentally. This is just the academic value of this manuscript prediction.

　　Otherwise, reader¨s ＾zero temperature￣ is a concept of nuclear structure, which means nucleons only occupy the lowest energy levels in order without a distribution among all energy levels, not mean zero energy including zero kinetic energy as the zero temperature in thermodynamics, of the order of 1-50 MeV for nucleons moving in the mean field. It is different from the meaning of about zero temperature (ultracold) corresponding to about zero energy in this manuscript at all.
\end{description}

\item
\begin{description}
\item[Ask] In contrast to the author's statement "We think the nuclear theory is not good enough to confirm this prediction or conclusion but experiments could confirm it" enough is known about pairing energies and scattering phase shifts to verify that a bound state is not possible, and the weak attraction of forming a Cooper pair would be insufficient to form a bound state.

Admittedly, it is believed that neutrons both in the crust and cores of neutron stars form superfluids in either the 1S0 or 3P2 state, but the gas of neutrons is confined by the large pressures in the neutron star interior. Even so, the energy change due to formation of Cooper pairs is negligible compared to their kinetic energy.
     \item[Answer] Here reader applied nuclear structure approaches to this manuscript conditions without paying attention to the basic physical assumptions used both in energy region and nuclear force. ＾pairing energies and scattering phase shifts￣ are concepts used in mean field theory which assumes basicly an independent nucleon moving in the mean field formed by the rest neucleons with energy 1-50 MeV, or reacting with incoming energy of order of MeV. Mean field approaches work well for stable heavy nuclei, not well for light nuclei, not work for nuclei with a few nucleons, can not describe the loosely binding halo nuclides near the drip line. There is no scattering phase shifts in the $10^{-7}$ eV or lower energy region and at the boundary region of nuclear force.

　　But here it is two free neutrons before forming a neutron pair, there is no mean field at all for a free neutron. Two free neutrons form a pair with the whole nuclear force which is unclearly now, especially in the boundary region of nuclear force, not only with the pairing residue interaction.

　　The key point for a bound state is the separate energy which reader omitted and did not indicate the ＾good enough￣ nuclear theory can confirm the separate energy to the accuracy of the order of MeV or keV, or even eV like those drip-line nuclei needed zero separate energy. Here the incoming interacting free neutrons with kinetic energy of $10^{-7}$ eV, both nuclear force/potential and nuclear theory is not good enough for it. So no evidence to support reader¨s conclusion that nuclear theory good enough can verify the possibility of any bound state including keV or eV or lower separate energy.

　　In nuclear astrophysics fields, they use the same nuclear theory as in nuclear structure or nuclear reaction. So the basic physical assumptions are the same as above and so do the conclusions.
\end{description}

\item
\begin{description}
\item[Ask] There is no reference to the literature concerning dineutrons, nor is there an explicit experimental path suggested to create these pairs.
     \item[Answer] this manuscript is the first to predict a possible way to create a new type of matter, free-neutron Bose-Einstein condensate by the ultracold free-neutron-pair Bose-Einstein condensation and then determine the neutron drip line experimentally. It is natural that there is no literature in this field.

　　In this manuscript there is an explicit experimental path suggested to create neutron pairs, i.e. medium atoms are cooled by the laser, then to cool the cold neutrons with the colder medium atoms to the super low temperature, trap ultracold neutrons, then to cool them, force or select higher energy atoms and neutrons to evaporate to make the rest neutrons colder reach to or below critical temperature or Fermi temperature, and become the high degenerate quantum Fermi gas, then form neutron pairs. One possible mechanism or way to form the pairs is by the strong interaction between two neutrons near the interacting boundary of nuclear force, or by the magnetic interaction strengthened by magnetic resonances between the two interacting neutrons.
\end{description}

\item
\begin{description}
\item[Ask] The argument seems based on the fact that deuterium exists, which is irrelevant to the present situation.
\item[Answer] In this manuscript deuterium is
　　(1) an example to show that two neutrons form a very loose binding boson not violating first principles, e.g. uncertainty principle, Pauli principle, etc.
　　(2) to show the binding energy or separate energy of one nucleon is quit different, deuterium about 2 MeV and 8 MeV for heavy nuclei.

　　One argument is the property of nuclear force, i.e. the long distance part of nuclear force is attractive and the short part is repulsive, which is confirmed and demonstrated by experiments. So it is attractive interaction near the boundary of nuclear force as stated in this manuscript.
\end{description}

\item
\begin{description}
\item[Ask] The manuscript does provide any evidence for any resonance states which would be necessary to stabilize this configuration even for a short time.
     \item[Answer] Here reader¨s resonance states, not clearly, should be those in nuclear reactions with energy of MeV, I think. In this manuscript the incoming neutron energy is of $10^{-7}$ eV or lower and interacting between two neutrons near the interacting boundary of nuclear force, for serious calculations, you would like to suggest based on QCD or nuclear force / potential, but no nuclear force derived from first principles now so your suggestion to confirm it with theory is impossible to perform by the field, I think.

　　So we can not give the formula in this boundary region of nuclear force and can not confirm it with the existing nuclear force. That is why we suggest to confirm it with experiments only.
\end{description}

\item
\begin{description}
\item[Ask] I do not find sufficient merit in this paper to warrant publication. Many reasons were given by at least one reader to support a rejection, and I will not reiterate them here although I find them all valid.
\item[Answer] First of all, as a strict scholar, one should check/verify the basic physical assumptions of a model or theory, and the range it works well because no any model or theory can work well in all conditions. Then to determine what kind of theory/approach is suitable for the physical phenomena in this manuscript.
　　This is the key step for readers and reader, but they all omitted.

　　I think reader¨s comments is unfair and incorrect, because
　　(1) reader only accepted the reader¨s judgements and comments which are based on the wrong basic assumptions both in energy region and nuclear force that general nuclear theory works well, did not seriously consider author¨s respondences to the readers and almost did not accept it.

　　By the reader¨s comments given, no anywhere to mention the two key points, i.e. energy region and nuclear force where nuclear theory works well, and the opposite opinions in the content of the author¨s respondence to the three readers.

　　 (2) reader gave many conclusions, without showing any theory or models used and their suitable region in which the models or theory works well and the basic assumptions used for the theory or models, without showing any actual evidence to support reader¨s conclusions too. But it is impossible for reader to be based on first principles, since there existing no nuclear force derives from first principles.

　　So I think reader¨s conclusions are based on the wrong assumptions both in energy region and nuclear force as the readers did and are incorrect, and unfair too.

　　(3) Fairly state, for reader, especially to the opposite results shown by readers and authors, should consider author¨s respondence very carefully and seriously since, compare to readers, the author considers the same question much deeper and take much longer time, even for many years as what I do. But here reader did not do so and showed an unfair hearing seriously.

　　THIS MANUSCRIPT content concerning many quit different fields, include: a) the Bose-Einstein condensation of bosonic and fermionic atoms in atomic gases (BEC) experiment, cooled by the laser, trap, evaporate cooling, forming pair or molecular for fermionic atoms, BEC detect, etc. b) nuclear structure both theory and experiment. c) nuclear reaction both theory and experiment. d) cold neutron, ultracold neutron experiment and its techniques.

　　Frankly speaking, unless the readers and reader are already an expert in all these fields listing above, if not, they only took a few weeks from receiving the manuscript this manuscript to sending out the comment report, it is very difficult to family with all these fields and it is impossible to become an expert in all these fields in such short time, since this manuscript has taken the author about 9 years, 2005-2013, in all these fields to perform it.

　　To conclude firmly that something is impossible in the field, one at least is an expert in the same field to warrant one¨s knowledge and understanding deep enough in that field. Otherwise his conclusion would be unreliable or even incorrect because to disconfirm one thing needs more knowledge and deeper understanding than favor it in the same field.

　　So I think readers  should consider the author¨s respondence seriously and carefully because compare to readers, the author considers the same question much deeper and take much longer time.

　　(4) I think the manuscript this manuscript has very high academic values because so many innovations were predicted or suggested. And the three readers have not reviewed it seriously and correctly. They judge this manuscript only by their assumptions not by the assumptions in this manuscript which is on first principles and different from their assumptions. They have given the incorrect comments based on the wrong assumptions, used the incorrect theory and methods both in energy region and nuclear force. So their conclusions are wrong and unreliable.

　　The manuscript this manuscript is based on first principles, not on approximate nuclear theory or models in which the accuracy is only about 1 MeV.

　　It is well known that in different energy region different theory or method should be used in nuclear physics. What the readers used is the general nuclear theory which is suitable in the 1--50 MeV energy region or higher. But in this manuscript the energy is about $10^{-7}$ eV. Here the difference is about 10--13 order of magnitude in the two situations. They used the same methods for such low energy incoming neutrons and must give the incorrect conclusions.

　　The other is nuclear force that readers think true correction or truth. The real situation is the existing nuclear force can not describe the loosely binding halo nuclides near the drip line and not derive from first principles. It is unclearly now, especially in the boundary region of nuclear force. So we can not give the formula in this boundary region and can not confirm it with the existing nuclear force. That is why we suggest to confirm it with experiments only.
\end{description}

\item
\begin{description}
\item[Ask] Accept only papers that are scientifically sound, important to the field, and contain significant
new results in physics. We judge that these acceptance criteria are
not met by your manuscript.
\item[Answer] But I do not agree your conclusion and think it is unfair and incorrect because it is based on the incorrect assumptions and theoretical models. The key points include two aspects, i.e. nuclear force and work well energy region. I state my reasons and evidences below:

(1) First of all, when you try to apply a nuclear theoretical model to research a nuclear phenomenon, you should check the basic assumptions and suitable region of this model or theory to judge if it can work well in the special condition of the phenomenon. It is well known that no nuclear theoretical model can work well in all energy regions and all conditions since nuclear force is not as clear as electromagnetic force. Nuclear force is not completely clear now, especially near the boundary region of nuclear force. Similarly, experimental results in a given energy region can not confirm the same experimental results must be observed in a quite different energy region. What your editors have done on this manuscript conclusion, the problem is just lost this important step.

(2) Work well energy region. The evidence you used, the negative neutron-neutron scattering length has been measured, are measured experimentally in the about 1-50 MeV or above energy region not in the $10^{-7}$ eV energy region. Here the energy difference is about 10-13 order of magnitude in the two situations. This negative scattering length can confirm only in the experimental energy region but it can not confirm and not deduce that it is negative in all energy regions. You and the other editors joined the discussion may check all your references involving the incident neutron energy of all neutron scattering experiments for neutron-neutron scattering length, and I believe you can not find any one in which incident neutron energy is in the $10^{-7}$ eV energy region or below. Hence, you use the results about 10！13 order of magnitude higher in energy to deduce and conclude the results in the $10^{-7}$ eV energy region, it is not reasonable and no scientific evidence too, so your conclusions must be incorrect and unreliable.

(3) The other argument is nuclear force that you would think true correction or truth and the real situation is the existing nuclear force can not describe the loosely binding halo nuclides near the drip line and not derive from first principles. It is unclearly now, especially in the boundary region of nuclear force. So we can not give the formula in this region and can not confirm it with the existing nuclear force. That is why we suggest to confirm it with experiments only. Nuclear force is the basic difficulty in nuclear physics. The research of basic nuclear force is beyond the nuclear physics research field and becomes the main field of particle physics which can not offer the exact nuclear force formula based on first principles like the electromagnetic force by now.
(4) The manuscript is based on first principles, not on approximate nuclear theory or models in which the accuracy is only about 1 MeV. The existing nuclear experimental and theoretical results in the MeV energy region can not be used to confirm and conclude the very weak binding case results in the $10^{-7}$ eV or lower energy region, i.e. must be in their suitably usable region without over their basic assumptions and limitations, which is extremely important.

     The manuscript Predicts or suggests:
(a) a new type of matter, free-neutron Bose-Einstein condensate;
(b) its creation way;
(c) a new neutron cooling method: by other colder neutral atoms which are cooled by the laser;
(d) the mechanism of free-neutron-pair formation: One possible way to form the pairs is by the strong interaction between two neutrons near the interacting boundary of nuclear force; Other possible mechanisms to form the pairs are by the magnetic interaction strengthened by magnetic resonances due to magnetic moment of neutrons.
(e) potential methods to synthesize heavy nuclides on the neutron drip line and to determine the neutron drip line and whether there are long-life nuclide or isomer islands beyond the neutron drip line experimentally;
(f) possible way to produce macroscopic quantitative man-made neutron condensed matter.
\end{description}

\item
\begin{description}
\item[Ask] There is no experimentally observed bound di-neutron.
\item[Answer] The correct state should be: by now there is no experimentally observed bound di-neutron and no theoretical prediction for it in the about 1-50 MeV energy region in which experiments and theory have studied already. That is just the scientific value of this manuscript prediction is. Just the reverse, if the bound di-neutrons have been observed experimentally, this manuscript prediction is meaningless indeed. One similar example is that before the two ultracold fermionic atoms to couple into a boson is performed experimentally in the Bose-Einstein condensation there is no existing such bosons formed by two fermionic atoms at ordinary temperature, which can be found by experiment.
\end{description}

\item
\begin{description}
\item[Ask] No theoretical evidence.
\item[Answer] One argument is the property of nuclear force, i.e. the long distance part of nuclear force is attractive and the short part is repulsive, which is confirmed and demonstrated by experiments. So it is attractive interaction near the boundary of nuclear force as stated in this manuscript.
\end{description}

\item
\begin{description}
\item[Ask] I read the paper ＾possible way to create´￣where the author suggests possible paths to make Neutron BEC. While theoretically this path might be possible,

I am afraid that experimentally will be quite challenging and to me impossible. However, the discussion and the goal of the paper is not suitable in my opinion
for publication. I rather suggest the author to submit it to a more specialized journal such as Nuclear Instruments and Methods or something similar.
     \item[Answer] By the comments given, without pointing out any actual incorrectness, directly reaching to rejection conclusion, not a suitable academic comment, I think you has not read the manuscript this manuscript seriously or carefully and understood it correctly.

　　I do not agree your judgements for this manuscript because your reject conclusions, without any actual scientific evidence, without pointing out any actual error, are unfair, incorrect and unreliable.

　　The discussion and the goal of the paper this manuscript, I state them below:

　　The manuscript this manuscript is based on first principles, not on approximate nuclear theory or models in which the accuracy is only about 1 MeV.

　　This manuscript focuses on nuclear physics mechanism. Actually, in this manuscript, almost each step is seriously related to the key nuclear physics issue, i.e. nuclear force. First step, free neutrons are cooled by the colder atoms and it is in nature the nuclear procedure of nuclei of atoms interacting with neutrons with nuclear force, or neutron-nucleus elastic scattering, the typical and main nuclear physics issue.

　　Other nuclear physics issue. For examples, one way to determine the neutron drip line, to synthesize very neutron-rich nuclides near or on the neutron drip line, to produce macroscopic quantitative man-made neutron condensed matter, all of these are nuclear physics hot issue which are predicted by this manuscript and can be performed by experiment. We think all these issues should be suitable and good.

　　It is well known that in different energy region different theory or method should be used in nuclear physics. The general nuclear theory is suitable in the 1-50 MeV energy region or higher. But in this manuscript the energy is about $10^{-7}$ eV. Here the difference is about 10-13 order of magnitude in the two situations. It is new energy region and a new research field indeed.

　　It is very difficult to perform the suggested goal in this manuscript experimentally should not be the reason to reject it. What you have done to reject it indicate your conclusions are unfair and incorrect.

　　The other is nuclear force that you think true correction or truth. The real situation is the existing nuclear force can not describe the loosely binding halo nuclides near the drip line and not derive from first principles. It is unclearly now, especially in the boundary region of nuclear force. So we can not give the formula in this region and can not confirm it with the existing nuclear force. That is why we suggest to confirm it with experiments only.

　　The manuscript this manuscript Predicts or suggests:
(1) a new type of matter, free-neutron Bose-Einstein condensate;
(2) its creation way;
(3) a new neutron cooling method: by other colder neutral atoms which are cooled by the laser;
(4) the mechanism of free-neutron-pair formation: One possible way to form the pairs is by the strong interaction between two neutrons near the interacting boundary of nuclear force; Other possible mechanisms to form the pairs are by the magnetic interaction strengthened by magnetic resonances due to magnetic moment of neutrons.
(5) potential methods to synthesize heavy nuclides on the neutron drip line and to determine the neutron drip line and whether there are long-life nuclide or isomer islands beyond the neutron drip line experimentally;
(6) possible way to produce macroscopic quantitative man-made neutron condensed matter.

　　Consider the list above, I think your comment " not suitable in my opinion for publication " is unfair and incorrect to this manuscript.

　　In summary I think the manuscript this manuscript has very high academic values because so many innovations are predicted or suggested. I think all these issues should be suitable and very good.
\end{description}

\item
\begin{description}
\item[Ask] This paper proposes techniques to trap and cool neutrons at sizeable
nuclear densities below the critical temperature for superfluidity. What a
dream!

The authors spend on the feasibility of their method no more than a few
sentences: cooling of neutrons by other colder atoms. These explanations
are given essentially in 3rd and 4th paragraph on page 4. For a proposal
of these potentially far reaching consequences, this is much too short
and, thus, hypothetical. No details or estimates are given how the
proposed cooling and trapping shall proceed. Also the way how the neutron
gas shall be compressed to nuclear densities is only very vaguely
sketched. I am not an experimentalist but, in my belief, the proposal to
make sense must contain a precise estimate of its realisability. In order
to compress a neutron gas from atomic gas densities to nuclear densities,
it seems to me that also the oscillator length of the trap must become
smaller by several orders of magnitude. I have no idea how this can be
achieved.

It seems to me that there is also a misconception on the theoretical side.
There is no BEC of neutrons because neutrons have no bound state in free
space. Neutrons also do not get bound within a gas of other neutrons, that
is chemical potential never becomes negative, see e.g. Matsuo et al. On
the other hand a gas of neutrons is always superfluid in the sense of weak
coupling BCS theory. But why to insist so much on superfluidity? A gas of
non-superfluid neutrons at typical nuclear densities trapped and kept for
measurable times would already be spectacular!

In conclusion the paper is not acceptable for publication. I strongly
doubt that revisions will change anything about this fact.
     \item[Answer] As what you understand, the goal of this manuscript is great indeed. So it has a very high academic value.

　　The style of this manuscript is a letter. In such a short paper, we think it should be as concise as possible, as the letter required.

　　The cooling and trapping technology, we think, is mature technology developed in BEC and cold neutron field, so we only point out the method and think it is enough to understand the whole procedure and mechanism. For the details one can get it in the references of this manuscript.

　　The key step of the feasibility of this method is to form neutron-pair as we have point out in this manuscript but readers omitted it. We have shown the mechanism and method needed, i.e. two ultracold neutrons couple into a pair with spin zero with nuclear force near it boundary region, then form free-neutron-pair BEC.

　　Reader¨s assumption in the comment on the density of free-neutron-pair BEC should be the same as that of nuclei is unreasonable and without any evidence to support this. For example, the density of BEC of atoms is about 10 orders of magnitude lower than that of its liquid or solid state. So the density of free-neutron-pair BEC should be very much lower than that of nuclei because it is a new type of matter predicted in this manuscript.

　　What this manuscript hopes or wants to solve is the whole mechanism or principle. For the actual technique to perform the goal in this manuscript would be developed further. It is not the purpose of this manuscript and to require in one letter to solve all problems are unreasonable.

　　For ＾neutrons have no bound state in free space￣.

　　First of all, when you try to apply a nuclear theoretical model to research a nuclear phenomenon, you should check the basic assumptions and suitable region of this model or theory to judge if it can work well in the special condition of the phenomenon. It is well known that no nuclear theoretical model can work well in all energy regions and all conditions since nuclear force is not as clear as electromagnetic force. Nuclear force is not completely clear now, especially near the boundary region of nuclear force. Similarly, experimental results in a given energy region can not confirm the same experimental results must be observed in a quite different energy region. What readers have done on the this manuscript conclusion, the problem is just lost this important step.

　　The evidence you used, the positive chemical potential, or the negative neutron-neutron scattering length has been measured, are measured experimentally in the about 1-50 MeV or above energy region not in the $10^{-7}$ eV energy region. Here the energy difference is about 10-13 order of magnitude in the two situations. This positive chemical potential or negative scattering length can confirm only in the experimental energy region but it can not confirm and not deduce that it is positive or negative in all energy regions. You may check all your references involving the incident neutron energy of all neutron scattering experiments for neutron-neutron scattering length, and I believe you can not find any one in which incident neutron energy is in the $10^{-7}$ eV energy region or below. Hence, you use the results about 10-13 order of magnitude higher in energy to deduce and conclude the results in the $10^{-7}$ eV energy region, it is not reasonable and no scientific evidence too, so your conclusions must be incorrect and unreliable.

　　Here readers applied nuclear structure mean field approaches to this manuscript conditions without paying attention to the basic physical assumptions used both in energy region and nuclear force. Neutron superfluid in BCS theory is concepts used in mean field theory which assumes basicly an independent nucleon moving in the mean field formed by the rest neucleons with energy 1-50 MeV, or reacting with incoming energy of order of MeV. Mean field approaches work well for stable heavy nuclei, not well for light nuclei, not work for nuclei with a few nucleons, can not describe the loosely binding halo nuclides near the drip line.

　　But here it is two free neutrons before forming a neutron pair, there is no mean field at all for a free neutron. Two free neutrons form a pair with the whole nuclear force which is unclearly now, especially in the boundary region of nuclear force, not only with the pairing residue interaction. So BCS theory and mean field approach are not suitable at all for free neutrons, and for a two-neutron-pair.

　　The manuscript this manuscript Predicts or suggests:
(1) a new type of matter, free-neutron Bose-Einstein condensate;
(2) its creation way;
(3) a new neutron cooling method: by other colder neutral atoms which are cooled by the laser;
(4) the mechanism of free-neutron-pair formation: One possible way to form the pairs is by the strong interaction between two neutrons near the interacting boundary of nuclear force; Other possible mechanisms to form the pairs are by the magnetic interaction strengthened by magnetic resonances due to magnetic moment of neutrons.
(5) potential methods to synthesize heavy nuclides on the neutron drip line and to determine the neutron drip line and whether there are long-life nuclide or isomer islands beyond the neutron drip line experimentally;
(6) possible way to produce macroscopic quantitative man-made neutron condensed matter.

　　In summary I think this manuscript has very high academic values because so many innovations are predicted or suggested. Reader¨s judgements and conclusions for this manuscript are incorrect on the most basic points, are unfair and unreliable too.
\end{description}

\end{enumerate}


\begin{thebibliography}{99}

\bibitem{And95} M. H.\ Anderson {\it et al.}, Science {\bf 269}, 198-201 (1995).

\bibitem{Dav95} K. B. Davis {\it et al.}, Phys. Rev. Lett. {\bf 75}, 3969-3973 (1995).

\bibitem{Reg03} C. A. Regal {\it et al.}, Nature {\bf 424}, 47-50 (2003).

\bibitem{Gre03}  M. Greiner {\it et al.}, Nature {\bf 426}, 537-540 (2003).

\bibitem{Ein25}  A. Einstein, Sitzungsber, Kgl. Preuss, Akad. Wiss. {\bf 1925}, 3 (1925).

\bibitem{Hu87} J. M. Hu {\it et al.}, Nuclear theory Vol. 1 (in
Chinese, Atomic Energy Press, Beijing, ed. 1, 1987), pp. 26-30.

\bibitem{Ign96}  V. K. Ignatovich, Physics--Uspekhi {\bf 39} (3), 283 -304 (1996).

\end{thebibliography}
\end{document}